\documentclass[11pt,a4paper]{article}
\usepackage{graphicx} 
\usepackage{jinstpub}
\usepackage{hyperref}
\usepackage{textcomp}
\usepackage{lineno}
\usepackage{ulem}

\usepackage{pifont}
\usepackage{color}

\title{A Simple Apparatus for Testing PMT Humidity Tolerance}
\author[1]{A.~Germer%
\note{Presently at Department of Physics \& Astronomy, University of Pennsylvania, Philadelphia, PA 19104.},}
\author{K.~Park,}
\author{C.~Skuse,}
\author{C.~Yang,}
\author[2]{and D.~S.~Parno%
\note{Corresponding author.}}

\affiliation{Department of Physics, Carnegie Mellon University,
Pittsburgh, PA, USA 15213}
\emailAdd{dparno@cmu.edu}
\date{March 2025}

\abstract{
We report on a low-cost apparatus to extend a  photomultiplier tube (PMT) testing setup to operations at high humidity and/or at an elevated temperature. This setup allows a determination of whether a PMT can successfully operate for an extended period of time in a high-humidity environment, such as the waterline of a water Cherenkov detector. 
}


\begin{document}

\maketitle

\section{Introduction}

Photomultiplier tubes (PMTs) have been a vital, reliable tool for photon detection almost since the first combination of a photocathode with an electron amplification stage, 90 years ago~\cite{iams:1935}. Their applications range from basic physics to life science to industrial use, but their essential design is vulnerable to environmental factors such as humidity, magnetic fields, and temperature~\cite{pmthandbook}. Many experiments are able to carefully control the ambient atmosphere, but high-humidity conditions can be unavoidable in applications requiring light sensing outdoors or in extreme environments. For example, Tanaka and Sannomiya describe a muon tomography system, used to image landslides, in a tunnel with nearly 100\% humidity; in this case, the entire apparatus, including power supplies, shared high-humidity conditions~\cite{Tanaka:2013}. 

The primary risk of high humidity to PMT operation arises from the infiltration of moisture into the PMT base and electrical connections: the surface leakage current increases exponentially with humidity levels~\cite{printedcircuits-handbook}, and at high humidity sparking and circuit damage may occur. This risk can be mitigated by coating  circuits in silicone or a similar protectant. In applications requiring PMT immersion (e.g., PMTs below the waterline in a water Cherenkov detector), the entire electrical assembly can be potted in epoxy to make it watertight (see, e.g., Ref.~\cite{Suzuki:1992as, SNO:1999crp, Alimonti:1998nt, Milagro:1999qhb, An:2013mea, DayaBay:2015meu}). Some PMT models are vulnerable to an additional danger: in a humid environment their hygroscopic entrance windows will adsorb water layers~\cite{Pashley:1979} that will alter their optical properties.

This work was undertaken to determine whether Hamamatsu R7081 PMTs, purchased as spares for the Double Chooz experiment~\cite{DoubleChooz:2022ukr} and potted for immersion in mineral oil, could safely be operated at the waterline of a water Cherenkov detector. 
Previous qualification tests for the PMTs were performed under ambient environmental conditions~\cite{Bauer:2011ne}, and did not address operations in a high-humidity environment. 
The detector in question is the second module of a two-module D$_2$O detector~\cite{COHERENT:2021xhx} under development to benchmark the neutrino flux from Oak Ridge National Laboratory's Spallation Neutron Source, in order to reduce the flux uncertainty on cross-section measurements by the COHERENT collaboration~\cite{Akimov:2022oyb}. Due to space constraints at the deployment location, the PMTs are located only at the top of this detector, so their input windows are submerged while the dynode chains and other electronics lie above the waterline. Even above the waterline, however, the operational environment can be expected to be very high-humidity. 

In order to test the operability of this PMT model in high humidity, we constructed a low-cost test setup featuring a watertight box into which we injected high-humidity air, and which allowed slight elevation of temperature. Section~\ref{sec:env-chamber} describes this environmental chamber in detail, while Sec.~\ref{sec:pmt-tests} shows an example of test data acquired with the chamber. In Sec.~\ref{sec:discussion} we summarize cost and performance.

\section{Environmental chamber}
\label{sec:env-chamber}

\begin{figure}[tbp]
\centering\includegraphics[width=0.8\textwidth]{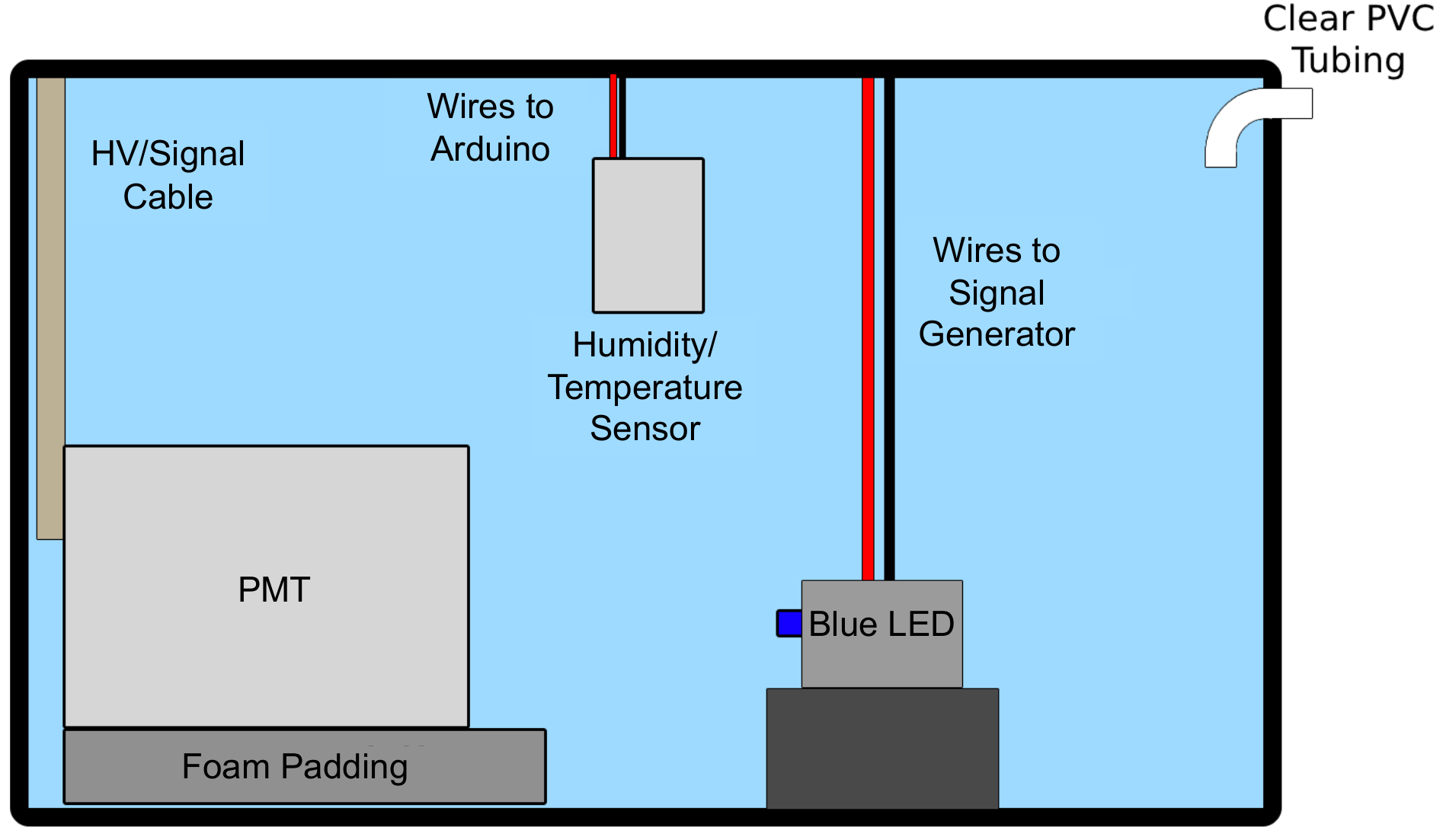}
\caption{A cutaway schematic of the plastic container that holds the PMT. Several cables are fed into the box through tight-fitting holes.}
\label{box_cutaway}
\end{figure}

\begin{figure}[tbp]
\centering
\includegraphics[width=0.8\textwidth]{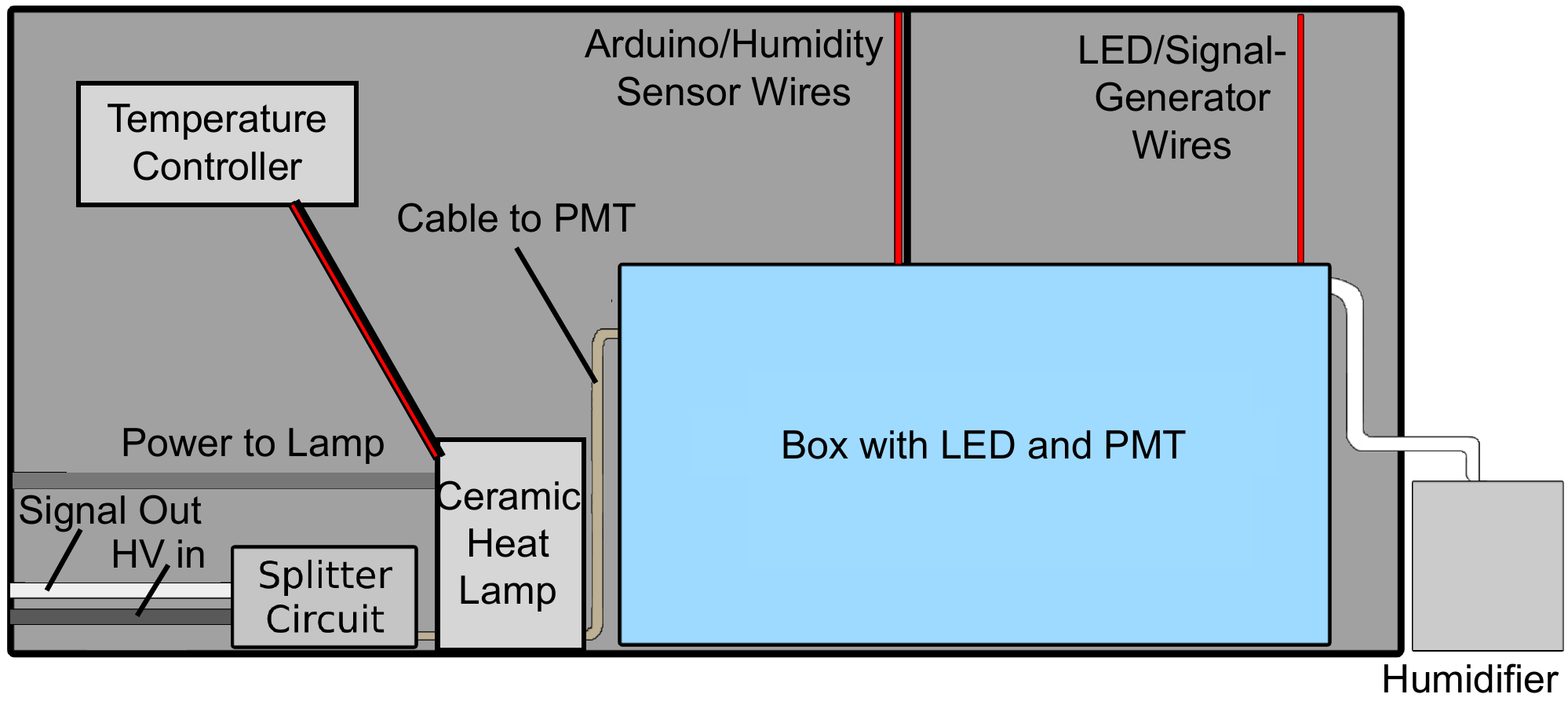}
\caption{A cutaway schematic of the entire setup. The large box is the dark box while the smaller box is the plastic container that holds the PMT. Cables are fed into the box through feedthrough connectors or tight-fighting holes.}
\label{whole_cutaway}
\end{figure}

Since PMTs are sensitive to visible light, a dark environment is essential for operations. Under normal environmental conditions, a wooden dark box is a common solution for testing. Our first technical challenge is to create a high-humidity environment inside the pre-existing, wooden dark box in our laboratory. We placed a plastic, 30-gallon HDX\textregistered{}~container inside the dark box to separate the high-humidity environment from vulnerable electronics and wood on the outside. Inside the plastic container was the PMT (cushioned on foam padding to prevent damage when the box was disturbed), a flashing blue LED, an Aosong DHT22/AM2302 temperature/humidity sensor, and a clear PVC tube feedthrough for introducing humidity. The porous foam padding is wrapped in plastic film to prevent penetration of moisture into slow-drying volumes conducive to mold growth. At very bright LED settings, reflections from the film become visible, and can be prevented with a layer of black fabric or paper. Figure~\ref{box_cutaway} shows a cutaway schematic of this waterproof volume, while Fig.~\ref{whole_cutaway} shows both nested boxes.

\begin{figure}[tbp]
\centering
\includegraphics[width=0.8\textwidth]{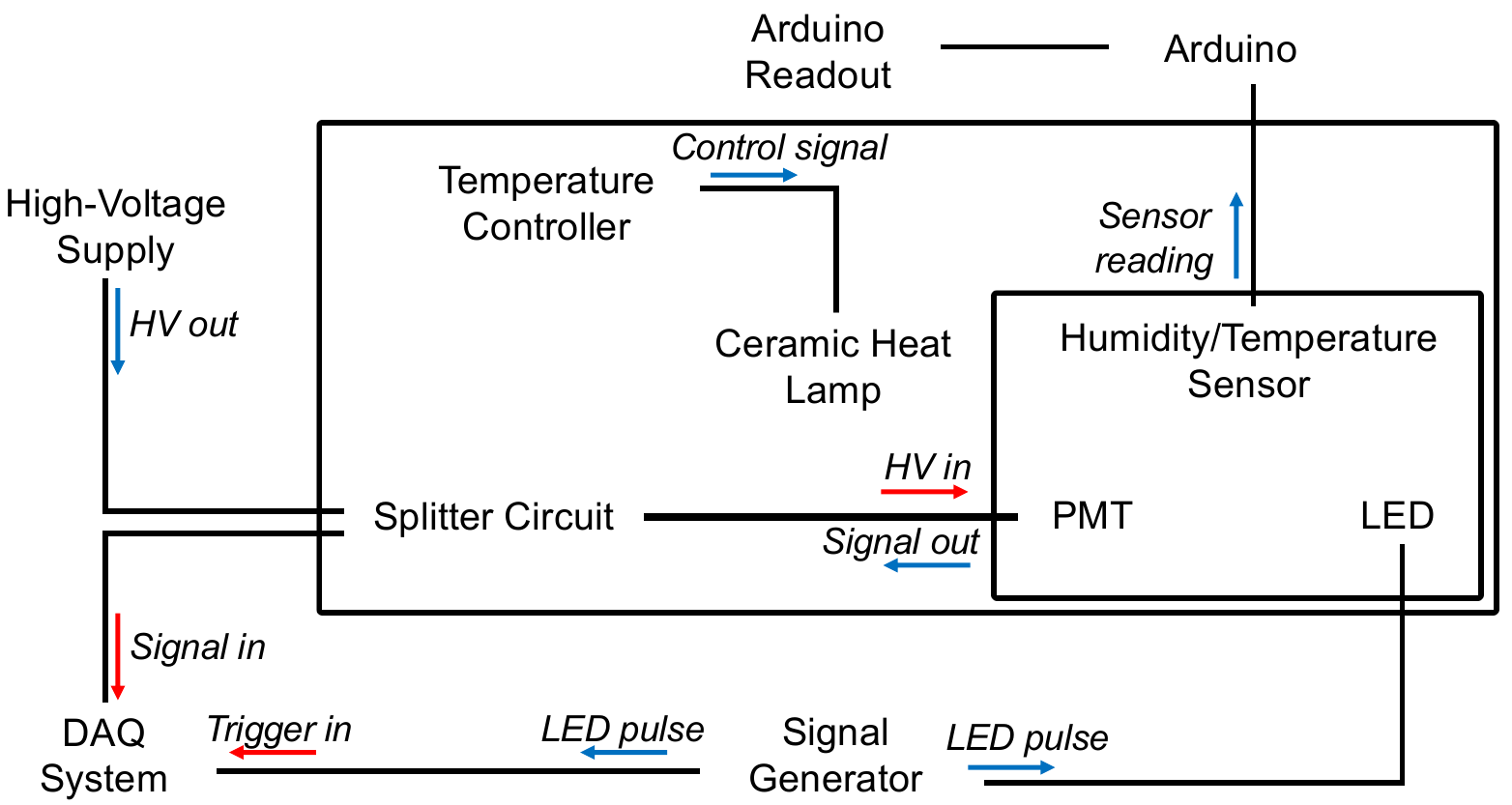}
\caption{Diagram of electronic connections in our setup. The larger box includes everything in the dark box and the smaller box includes everything inside the plastic container. The temperature controller includes a built-in temperature sensor.}
\label{box_diagram}
\end{figure}

Outside of the plastic container but inside the wooden dark box are a ceramic heating element and temperature controller (Sec.~\ref{sec:high-temp}) and a splitter circuit (Sec.~\ref{sec:pmt-light-daq}). Small feedthroughs admit cable connections to electronics outside both boxes. For example, the humidity/temperature sensor is connected to an ELEGOO UNO R3 Arduino board, read out on a Macbook Pro. Figure~\ref{box_diagram} diagrams the electrical connections. The calibration light source is a commercial blue LED (sold in a variety pack of 250 pieces) driven by a Global Specialties\textregistered{} model 4001 pulse generator located outside both the plastic container and the dark box. It is elevated on a small block for rough alignment with the PMT axis.

\subsection{Humidity injection}
\label{sec:humidity-inject}

To allow semi-continuous injection of humidity without exposing the test PMT to light, we place the humidity source -- either a Pure Enrichment\textregistered~1.5L humidifier or a Vicks\textregistered V4600~4.5L humidifier -- outside both boxes. A clear PVC tube carries the water vapor produced by the humidifier through tight-fitting holes in the dark box and plastic container.

\begin{figure}[ht]
\centering\includegraphics[width=0.8\textwidth]{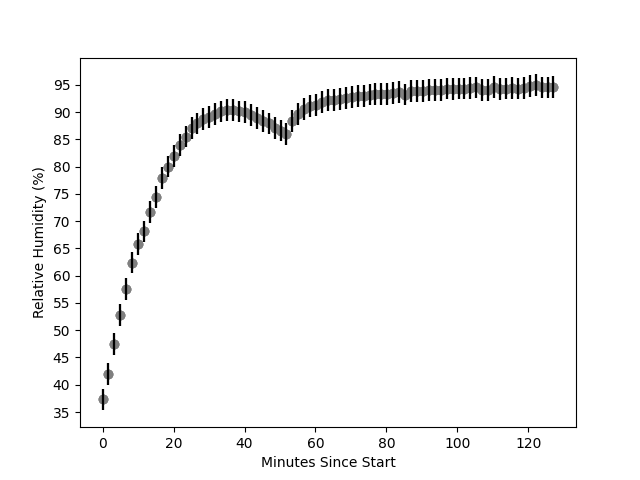}
\caption{A typical time series of humidity measurements with no PMT within the plastic container, taken by the humidity/temperature sensor in the plastic container and recorded by the Arduino. The Vicks humidifier is turned on at time $t=0$. A condensation blockage in the PVC tube was cleared at about minute 55.  The error bars show the manufacturer-specified relative accuracy of $\pm 2\%$. For orientation, the typical PMT run is one minute long.}
\label{fig:humidity}
\end{figure}

Figure~\ref{fig:humidity} shows the typical humidity evolution inside the plastic container during one of our runs\footnote{These data were acquired with the Vicks humidifier, but similar behavior is observed in both.}. We first collect data at room temperature; we then turn the humidifier on and wait for the humidity inside the plastic container to plateau. The process of bringing the environmental chamber up to high humidity is typically interrupted after about 15--60 minutes by the buildup of condensation inside the tubing, impeding the further flow of humid air. Condensation is removed by means of a pipe cleaner attached to a long piece of string on each end. The longer (3') string is threaded through the 2' plastic tubing and routed outside the environmental chamber. When the increase in relative humidity is observed to halt prematurely, this end of the string is pulled so that the pipe cleaner travels the whole length of the tube, absorbing moisture. If a second cleaning is necessary, pulling the shorter string (just over 2') brings the pipe cleaner through the tube in the opposite direction, toward the humidifier. 

The joints between the string and pipe cleaner are liable to fail after four or five uses, due to metal fatigue. This may be avoided by replacing the pipe cleaner after one round trip through the tube. To facilitate multiple cleanings, one might conceive replacing the pipe cleaner with a specially designed device, e.g. replaceable absorbent padding wrapping a thin rod with narrow eye hooks. 

During very long runs, e.g. the 5-day run described in Sec.~\ref{sec:pmt-tests}, condensation buildup becomes both less frequent and less predictable compared to the typical startup period shown in Fig.~\ref{fig:humidity}. With checks of the humidity level a few times a day and interventions as needed, the relative humidity level could be kept above about 70\%.

\subsection{Operation at elevated temperature}
\label{sec:high-temp}

To simulate a longer exposure to a high-humidity environment, we added a heating element to the inside of the dark box with the goal of increasing the moisture penetration rate into the PMT electronics. 
As shown in Fig.~\ref{whole_cutaway}, a Lucky Herp 150W ceramic heat emitter is placed at the bottom of the wooden box, outside the plastic container. The emitter is secured by a commercial heat-lamp base from Sheens. We use an Inkbird ITC-308 Digital Temperature Controller to toggle power to the heat emitter, based on a desired temperature of 30$^\circ$ C. These components are widely and cheaply available due to their use in temperature-control feedback loops in reptile cages.  

\begin{figure}[ht]
\centering\includegraphics[width=0.75\textwidth]{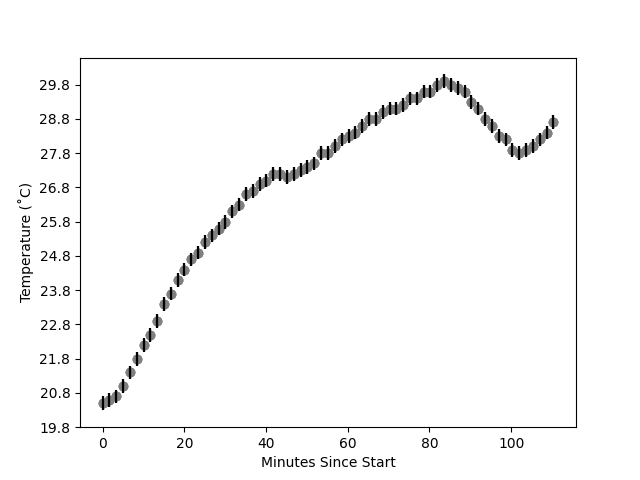}
\caption{A typical set of temperature measurements with no PMT within the plastic container, taken by the humidity/temperature sensor in the plastic container and recorded by the Arduino. The ceramic heat emitter is turned on at time $t=0$. Although the digital temperature controller (in the dark box) is set to 30$^\circ$ C, the humidity/temperature sensor in the nested plastic box typically reads slightly lower temperatures as shown. The error bars show the manufacturer-specified accuracy of $\pm 0.2^\circ$C.}
\label{temp_warmup}
\end{figure}

Figure~\ref{temp_warmup} shows the warmup period of the plastic box. The maximum temperature can be set via a keypad on the digital temperature controller, which consistently reported temperatures about 3.5$^\circ$C higher than that in the plastic box. Higher and more stable temperatures could have been achieved with a higher setpoint and the addition of insulation on the outer, wooden box. The manufacturer specified a maximum temperature of 50$^\circ$C for this model PMT. 

\subsection{Light sensing and data acquisition}
\label{sec:pmt-light-daq}

\begin{figure}[h]
\centering
\includegraphics[width=0.45\textwidth]{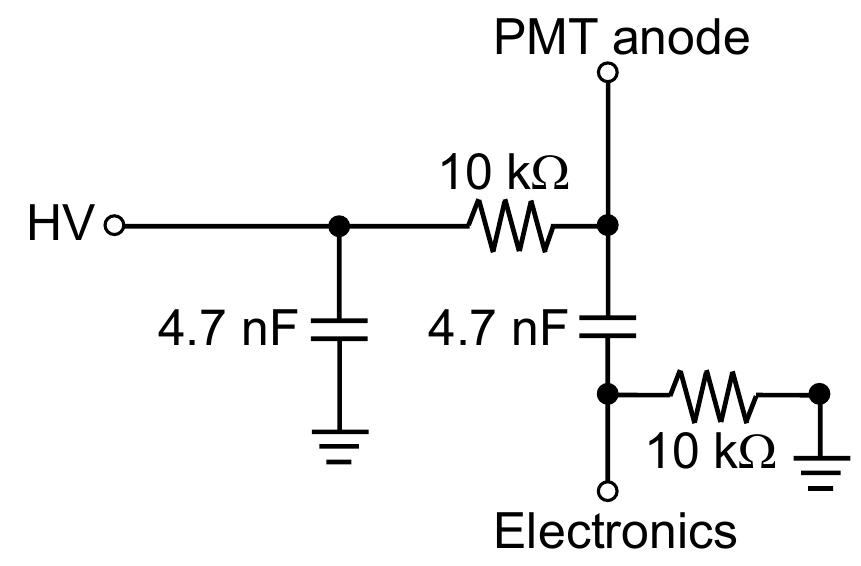}
\caption{Splitter circuit for decoupling the HV input and the PMT signal output to the data acquisition, labeled ``Electronics''.}
\label{splitter}
\end{figure}

In this work, we show the operation of this environmental chamber with a 10'' diameter Hamamatsu\textregistered{} R7081 PMT configured to operate with positive high voltage (HV)~\cite{Belver:2018erf}. In this configuration, the PMT base is connected to a single coaxial cable that carries both the positive high voltage (HV) and the PMT signal. A splitter circuit is needed to decouple the HV input and the signal output (see, e.g., Refs.~\cite{Bauer:2011ne, Sato:2012jsa, RENO:2010vlj}). The circuit deployed for our setup is shown in Fig.~\ref{splitter}, including a 10k$\Omega$ resistor to match impedance with the PMT base circuit~\cite{Bauer:2011ne}. The circuit is constructed inside a small, die-cast metal box placed outside the humidity box, and inside the dark box.

Data acquisition for our earliest tests was oscilloscope-based, similar to that used for the MicroBooNE test stand~\cite{TBriese_2013}. We used a Tektronix\textregistered{} model DPO7104 oscilloscope to write histogram and statistics data to a portable drive. Later, we reconfigured our primary data acquisition to use a CAEN 5720 desktop digitizer read out with WaveDump Software~\cite{CAEN-digi}. In both cases, event recording was internally triggered based on pulse height.

\section{PMT Testing in Environmental Chamber}
\label{sec:pmt-tests}

\begin{figure}[ht]
\centering\includegraphics[width=0.75\textwidth]{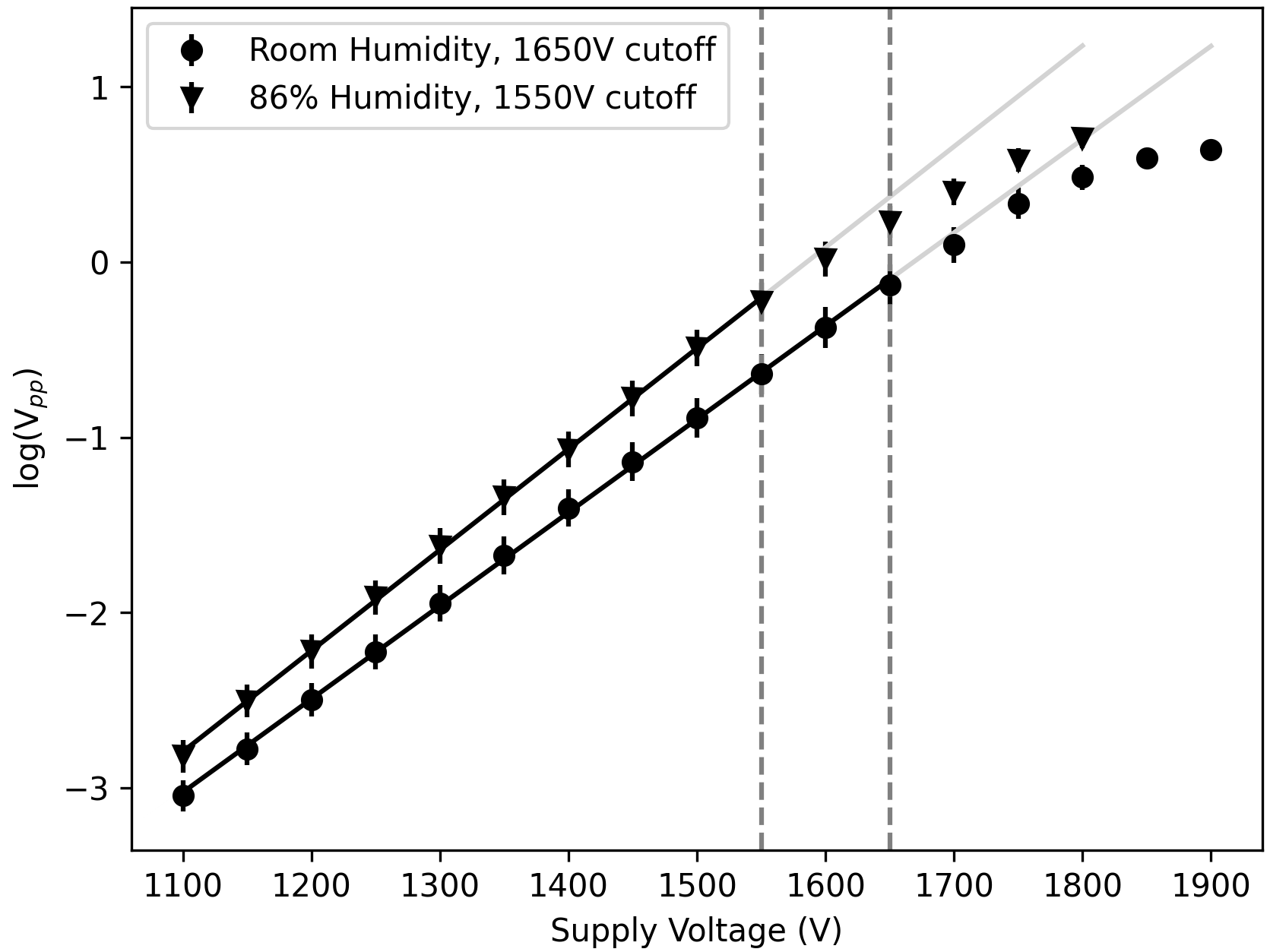}
\caption{PMT responses to a varied HV input, under 86$\%$ and ambient humidity conditions. For this run, the Pure~Enrichment\textregistered{} humidifier was used. Error bars are statistical only. The vertical dashed lines indicate observed ``cutoff'' voltages above which the assumption of a linear $\log V_{pp}$ response to applied high voltage begins to break down. Within these linear regions, the measured slope of $\log V_{pp}$ (with $V_{pp}$ in V) to supply voltage is $(5.32 \pm 0.03)\times 10^{-3}$/V for room humidity, and $(5.75 \pm 0.04)\times 10^{-3}$/V for 86\% humidity; each fit line is continued beyond the fit range in light gray. We attribute the offset between curves to a mechanical disturbance of the box between tests.}
\label{fig:HVhumidity}
\end{figure}

As an example of measurements taken with this apparatus, Fig.~\ref{fig:HVhumidity} shows the results of gain tests conducted at ambient and elevated humidity, with applied high voltage between 1100 and 1900~V. At each test point, we measure the amplitude $V_{pp}$ of the PMT signal, averaged over $\mathcal{O}(100,000)$ identically triggered LED pulses. We observe a linear $\log V_{pp}$ response for operating voltages up to 1550~V (86\% relative humidity) or up to 1650~V (room humidity).  The PMTs are rated to 2000V, with a ``typical" operating voltage of 1500V~\cite{pmtdatasheet}. We fit linear functions to each response curve below the corresponding cutoff voltage, and find the slopes to differ by 8\%, suggesting that the PMT gain curve must be calibrated at the operational humidity setting rather than being extrapolated from low-moisture conditions. Between HV sweeps, we found that the slopes remained unchanged whereas the y-intercept would vary. These changes in y-intercept are often correlated with a physical disturbance of the PMT setup, for example a slight shifting of the box, which may alter the internal alignment.

The PMT was also operated at a slightly elevated temperature. Once a stable temperature of about 26.5$^\circ$C was reached inside the plastic box under conditions of elevated humidity, the PMT's response to LED pulses was observed over the course of 5~days as a proof-of-concept measurement for long-term tests as described in Sec.~\ref{sec:high-temp}. No statistically significant change in the PMT response was observed.

\section{Discussion and conclusions}
\label{sec:discussion}

We have constructed a simple, low-cost chamber for testing PMT performance under conditions of elevated humidity and/or temperature. The chamber is an effective supplement to a standard bench setup for testing PMTs under ordinary environmental conditions. The total cost of the chamber, excluding labor and standard components for PMT testing (PMT, data acquisition, high-voltage supply, function generator, cables) is less than 200 US dollars (2022). Table~\ref{tab:costs} gives the cost breakdown of all components required for the chamber. 

\begin{table}[tbp]
   \centering
   \caption{Cost breakdown for a new environmental chamber in 2022 US dollars. Standard components for PMT testing (PMT, high-voltage supply with splitter circuit, data acquisition, function generator, cables, wooden dark box) are excluded. Our two humidifer models had the same price.}
    \label{tab:costs}
   \begin{tabular}{lcr} 
     \hline
    Component & Quantity Used & Price (\$ US)  \\
    \hline
    HDX\textregistered{} 30-gallon plastic container & 1  & 12.98 \\
    Foam padding (priced for 2''x24''x24'') & 2''x18''x19'' & 7.64 \\
    Plastic wrap (priced for 225 sq. ft.) & 6 sq. ft. & 4.00\\
    Arduino\textregistered{} UNO Rev3 board & 1 & 28.50  \\
    HiLetGo\textregistered{} DHT22/AM2302 humidity sensor & 1 & 7.00  \\
    Humidifier & 1 & 39.99 \\
    Vinyl tubing & 2 feet & 5.00 \\
    Pipe cleaner (priced for pack of 200) & 1 & 5.99 \\
    Nylon twine (priced for 250 feet) & 5 feet & 5.79 \\
    Blue LED (priced for pack of 250) & 1 & 4.99 \\
    Lucky Herp\textregistered{} 150W ceramic heat emitter & 1 & 11.99 \\
    Heating-element holder & 1 & 15.00 \\
    Inkbird\textregistered{} ITC-308 Digital Temperature Controller & 1 & 35.00 \\
    \hline
    Total & & \$183.87 \\
\end{tabular}
\end{table} 

The setup does not allow fine control of humidity levels, but as demonstrated in Fig.~\ref{fig:humidity}, can maintain a stable relative humidity in excess of 90\%. The temperature in the chamber, programmed using a commercial feedback controller, is stable to within a few degrees. The use of a box with a large aperture allows rapid drying of the interior and return to ambient humidity and temperature. 

Due to other operational considerations, the second module of the COHERENT heavy-water detector will use PMTs potted by the manufacturer for immersion in water, rather than the PMTs tested as part of this work. However, we conclude from the PMT performance demonstrated in this environmental chamber that these PMTs could be successfully operated at the waterline of a water Cherenkov detector, as initially hoped. The environmental-chamber design can be used to qualify PMTs (or other electronics) for operation under high humidity.

\acknowledgments
This work is supported by the U.S. Department of Energy, Office of Science, High Energy Physics, under Award No.\,DE-SC0022125. We gratefully acknowledge the loan of a PMT from Phil Barbeau, the loan of a digitizer from Hamish Robertson, and useful discussions with Eric Day, Yuri Efremenko, Gregg Franklin, Hexiang Huang, and Jon Link.


\providecommand{\href}[2]{#2}\begingroup\raggedright\endgroup

\end{document}